\documentclass[a4paper,10pt]{article}
\usepackage[utf8x]{inputenc}
\usepackage{graphicx}
\usepackage{amsmath}
\usepackage{cite}
\usepackage{authblk}
\usepackage{color}
\usepackage[margin=3cm]{geometry}
\PrerenderUnicode{ß}

\title{\bf Very early warning of a moderate-to-strong El~Ni\~no in 2023}
\date{}

\author[1] {Josef Ludescher}
\author[2] {Jun Meng}
\author[3] {Jingfang Fan}
\author[4] {Armin Bunde}
\author[1] {Hans Joachim Schellnhuber}

\tiny
\affil[1] {Potsdam Institute for Climate Impact Research, 14412 Potsdam, Germany}
\affil[2] {School of Science, Beijing University of Posts and Telecommunications, Beijing 100876, China}
\affil[3] {School of Systems Science, Beijing Normal University, 1000875 Beijing, China}
\affil[4] {Institute for Theoretical Physics, Justus Liebig Universit\"at Gießen, 35392 Gießen, Germany}

\begin{document}

\maketitle

\begin{abstract}
The El~Ni\~no Southern Oscillation (ENSO) is the strongest driver of year-to-year variations of the global climate and can lead to extreme weather conditions and disasters in various regions around the world.
Here, we review two different approaches for the early forecast of El~Ni\~no that we have developed recently: the climate network-based approach \cite{Ludescher2013} allows forecasting the onset of an El~Ni\~no event about 1 year ahead,
while the complexity-based approach \cite{Meng2019} allows additionally to estimate the magnitude of an upcoming El~Ni\~no event in the calendar year before.
For 2023, both approaches predict the onset of an El~Ni\~no event, with a combined onset probability of about 89\%. The complexity-based approach predicts a moderate-to-strong El~Ni\~no with a magnitude of $1.49\pm0.37$\textdegree C. Since El~Ni\~no events temporarily increase the global temperature, we expect that the coming El~Ni\~no will increase the global temperature by about +0.2\textdegree C, likely making 2024 the hottest year since the beginning of instrumental observations. It is possible that as a consequence of this El~Ni\~no, the +1.5°C target (compared to pre-industrial levels) will be temporarily breached already in 2024.
\end{abstract}

\section{The El~Ni\~no Southern Oscillation}

\begin{figure}[]
\begin{center}
\includegraphics[width=8cm]{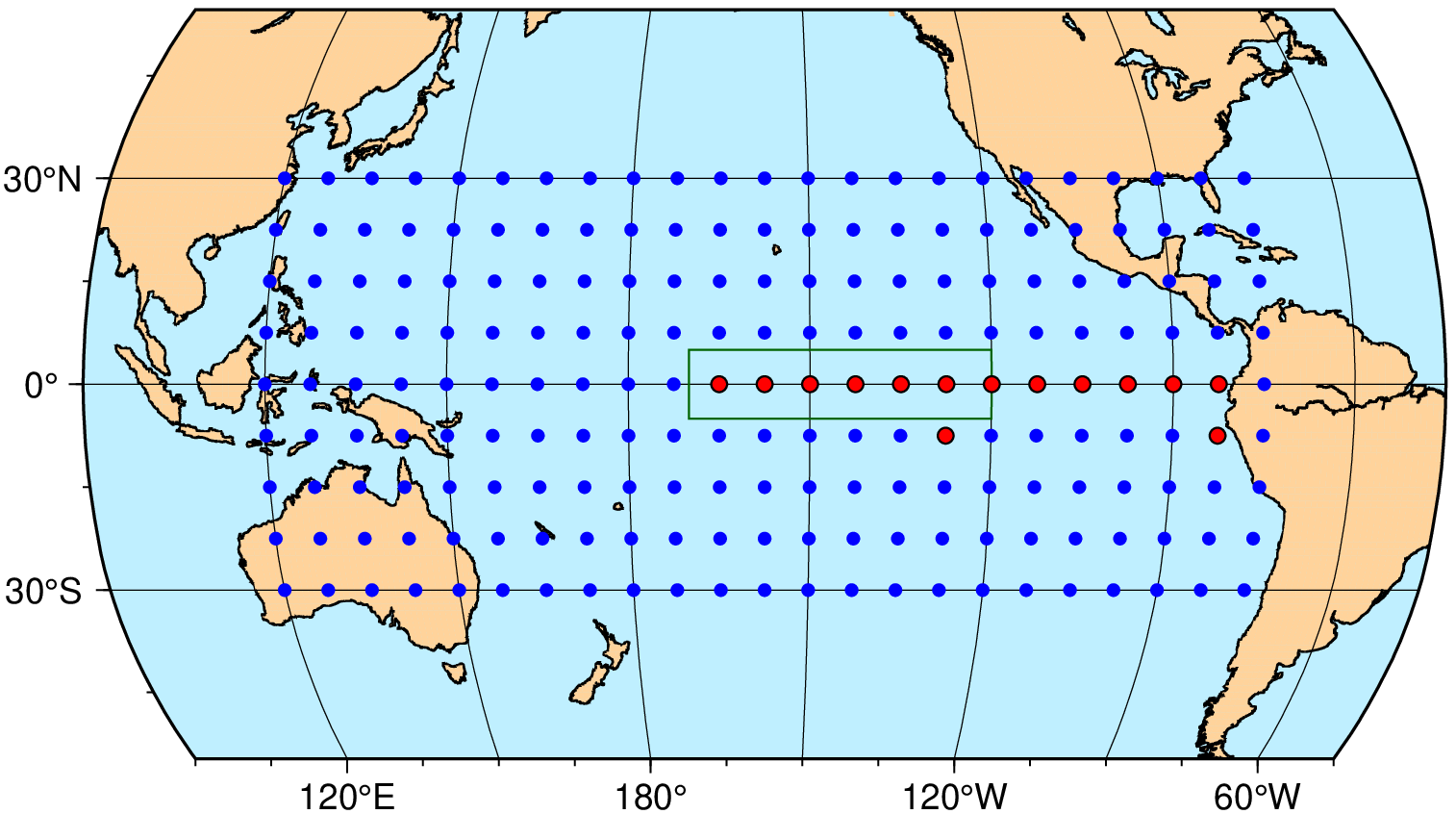}
\caption{The nodes of the climate network. The network consists of 14 grid points in the central and eastern equatorial Pacific (red dots) and 193 grid points outside this area (blue dots). The green rectangle shows the Ni\~no3.4 area. The grid points represent the nodes of the climate network that we use here to forecast the {\it onset} of an El~Ni\~no event. Each red node is linked to each blue node. The nodes are characterized by their surface air temperature (SAT), and the link strength between the nodes is determined from their cross-correlation (see below).}
\label{fig1}
\end{center}
\end{figure}

The El~Ni\~no-Southern Oscillation (ENSO) \cite{Clarke08, Sarachik10, Dijkstra2005, Wang2017,Timmermann2018, McPhadden2020} is a naturally occurring quasi-periodic oscillation of the Pacific ocean-atmosphere system that alternates between warm (El~Ni\~no), cold (La~Ni\~na) and neutral phases.
ENSO is quantified by the Oceanic Ni\~no Index (ONI), which is defined as the three-month running-mean sea surface temperature (SST) anomaly in the Ni\~no3.4 region (see Fig. \ref{fig1}). An El~Ni\~no episode is defined to occur when the ONI is greater or equal 0.5°C for at least 5 months. A regularly updated table of the ONI can be found at \cite{NOAA}.

Since El~Ni\~no episodes can alter weather conditions in various parts of the world and even lead to extreme weather conditions, like extreme precipitation or droughts \cite{Wen2002,Corral10,Donnelly07,Kovats03,Davis2001,McPhadden2020},
early-warning methods are highly desirable.
The two main types of prediction tools are dynamical and statistical models.
The dynamical models are initialized by observations and directly simulate the further development of physical quantities like the SSTA.
In contrast, the statistical models, which also include machine learning-based methods, exploit significant statistical relationships in current and past observations to obtain predictions.
Numerous models of both types have been proposed to forecast the pertinent index with lead times between 1 and 24 months
\cite{Cane86,Penland1995,Tziperman97,Kirtman03,Fedorov03,Galanti03,Chen04,Palmer2004,Luo08,Chen08,Chekroun11,Saha2014, Chapman2015, Lu2016, Feng2016,Rodriguez2016, Nootboom2018, Meng2018, Ham2019,DeCastro2020,Petersik2020,Hassanibesheli2022}.

Unfortunately, the current operational forecasts have quite limited anticipation power.
In particular, they generally fail to overcome the so-called ``spring barrier'' (see, e.g., \cite{Webster1995,Goddard2001}), which usually shortens their reliable warning time to around 6 months \cite{Barnston2012, McPhadden2020}.

To resolve this problem, we have recently introduced two alternative forecasting approaches \cite{Ludescher2013,Meng2019}, which considerably extend the probabilistic prediction horizon. The first approach \cite{Ludescher2013} (see also \cite{Ludescher2014,Ludescher2021a}) is based on complex-networks analysis \cite{Tsonis2006,Yamasaki2008,Donges2009,Gozolchiani2011,Dijkstra2019,Fan2020,Ludescher2021,Fan2022} and provides forecasts for the onset of an El~Ni\~no event, but not for its magnitude, in the year before the event starts.
The second approach \cite{Meng2019}
relies on the System Sample Entropy (SysSampEn), an information entropy, which measures the complexity (disorder) in the Ni\~no3.4 area. The method provides forecasts for both the onset and magnitude of an El Ni\~no event at the end of the previous year.

Here we present the forecasts of both methods for 2023. Both methods forecast the onset of an El~Ni\~no in 2023. There were 7 such concurring onset forecasts in the past, and all 7 turned out to be correct (see Figs. 2 and 5). Applying a conservative Bayesian-type estimate using Laplace's rule of succession \cite{Jaynes2003}, we arrive at an El~Ni\~no onset probability of $(7+1)/(7+2)=0.889 \approx89\%$.

Depending on the location of maximal warming, El~Ni\~no events can be divided into Central Pacific (CP) and Eastern Pacific (EP) events. The largest EP El~Ni\~no events are larger than the largest CP events. Indeed, between 1950 and present, the six largest El~Ni\~nos events were EP events. The type of an event itself also has a major influence on the event's impacts, see, e.g., \cite{Ashok2007, Weng2007, Wiedermann2021, McPhadden2020}.
For instance, large EP events typically lead to strongly increased precipitation along the coast of Ecuador and Northern Peru, resulting in massive floodings and landslides, while CP events only lead to dry conditions in these already dry areas  (see, e.g., \cite{Lagos2008,Bazo2013}). Based on \cite{Ludescher2022}, there is an 87.5\% probability that an El~Ni\~no event starting in 2023 will be an EP El~Ni\~no. The corresponding Bayesian estimate is 80\%.

\section{Climate network-based forecasting}

\subsection{The network-based forecasting algorithm}

The climate network-based approach is based on the observation that a large-scale cooperative mode, linking the central and eastern equatorial Pacific with the rest of the tropical Pacific (see Fig. 1), builds up in the calendar year before an El~Ni\~no event.
According to \cite{Gozolchiani2011,Ludescher2013,Ludescher2014}, a measure for the emerging cooperativity can be derived from the time evolution of the teleconnections (``links``) between the surface air temperature anomalies (SATA) at the grid points (''nodes``) between the two areas. The strengths of these links are derived from the respective cross-correlations (for details, see, e.g., \cite{Ludescher2013,Ludescher2014}).

The main predictive quantity for the onset of an El~Ni\~no is the mean link strength $S(t)$ in the considered network, which is obtained by averaging over all individual links at a given time $t$ \cite{Ludescher2013,Ludescher2014}.
The mean link strength $S(t)$ typically rises in the calendar year before an El~Ni\~no event starts and drops with the onset of the event (see Fig. 2). This property serves as a precursor for the event. The forecasting algorithm involves as only fit parameter a decision threshold $\Theta$, which has been fixed in a learning phase (1950-1980) \cite{Ludescher2013}. Optimal forecasts in the learning phase are obtained for $\Theta$ between 2.815 and 2.834 \cite{Ludescher2013, Ludescher2014}.

The algorithm gives an alarm and predicts the onset of an El~Ni\~no event in the following year when $S(t)$ crosses $\Theta$ from below while the most recent ONI value is below $0.5^\circ$C. In a more restrictive version (ii) \cite{Ludescher2022}, the algorithm considers only those alarms where the ONI remains below 0.5°C for the rest of the year.

For the calculation of $S$, we use daily surface air temperature data from the National Centers for Environmental Prediction/National Center for Atmospheric Research (NCEP/NCAR) Reanalysis I project \cite{reanalyis1,reanalyis2}. We would like to note that for the calculations in the prediction phase (1981-present), e.g., of the climatological average, only data from the past up to the prediction date have been considered.

\begin{figure}[h]
\begin{center}
\includegraphics[width=14cm]{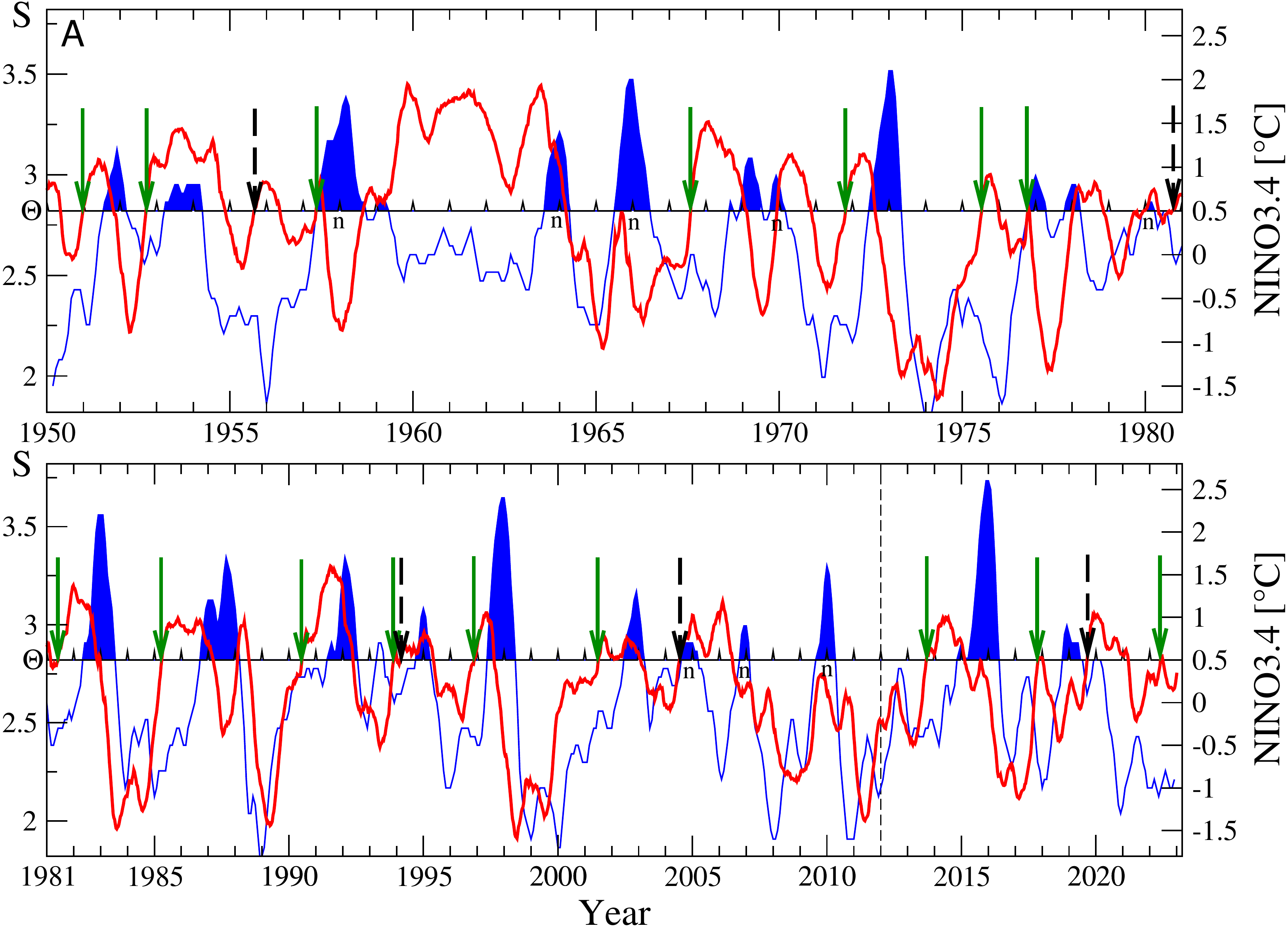}
\caption{The network-based forecasting scheme.  We compare the average link strength $S(t)$
in the climate network (red curve) with a decision threshold $\Theta$ (horizontal line, here $\Theta = 2.82$), (left scale), and the standard Ni\~no3.4 index (ONI), (right scale), between January 1950 and December 2022.
When the link strength crosses the threshold from below, and the last available ONI is below $0.5^\circ$C,
we give an alarm and predict that an El~Ni\~no episode will start in the following calendar year. 
The El~Ni\~no episodes (when the Ni\~no3.4 index is at or above $0.5^\circ$C for at least 5 months) are shown by the solid blue areas.
Correct predictions are marked by green arrows and false alarms by dashed arrows. The index $n$ marks not predicted events. The learning phase for the threshold is between 1950 and 1980. In the whole period between 1981 and December 2022, there were 11 El Ni\~no events. The algorithm generated 11 alarms and 8 of these were correct. In the more restrictive version (ii) of the algorithm, only those alarms are considered where the ONI remains below 0.5°C for the rest of the year. In this version the correct alarms in 1957 and 1976, and the incorrect alarms in 1994, 2004 and 2019 are not activated. Thus between 1981 and December 2022, version (ii) of the algorithm gives 8 alarms, all of which were correct.
}
\label{fig2}
\end{center}
\end{figure}

\subsection{El~Ni\~no forecasts since 2011}
The climate network-based algorithm has been quite successful in providing real-time forecasts, i.e., forecasts into the future. In its original version, it provided 11 forecasts for the period 2012-2022, 10 of these forecasts turned out to be correct, see Fig. 3.
The only incorrect forecast is a false alarm given in September 2019. The p-value, obtained from random guessing with the climatological El~Ni\~no onset probability, for the skill in the forecasting period alone is $p=0.017$. When considering the hindcasting and forecasting periods (1981-2022) together, the p-value is $p=3.6\cdot10^{-5}$. When version (ii) is applied, and only those alarms are considered, where the ONI remains below 0.5°C for the rest of the year, the false alarm in 2019 is not activated.

Between May 24th and June 7th, 2022, the mean link strength crossed all critical thresholds, while the last available ONI value was below $0.5$\textdegree C ($-1.1$\textdegree C for March-April-May), thus predicting the onset of an El~Ni\~no in 2023. In the hindcasting and forecasting period (1981 and December 2022), there were 11 El~Ni\~no events. The algorithm generated 11 alarms and 8 of these were correct. The more restrictive version (ii) of the algorithm \cite{Ludescher2022} gave 8 alarms, all of which were correct. Since the ONI remained negative throughout 2022, also the conditions for the more restrictive version of the network-based algorithm were met. Based on the climate network approach alone, a likely onset of an El~Ni\~no is thus predicted for 2023.

\begin{figure}[]
\begin{center}
\includegraphics[width=10cm]{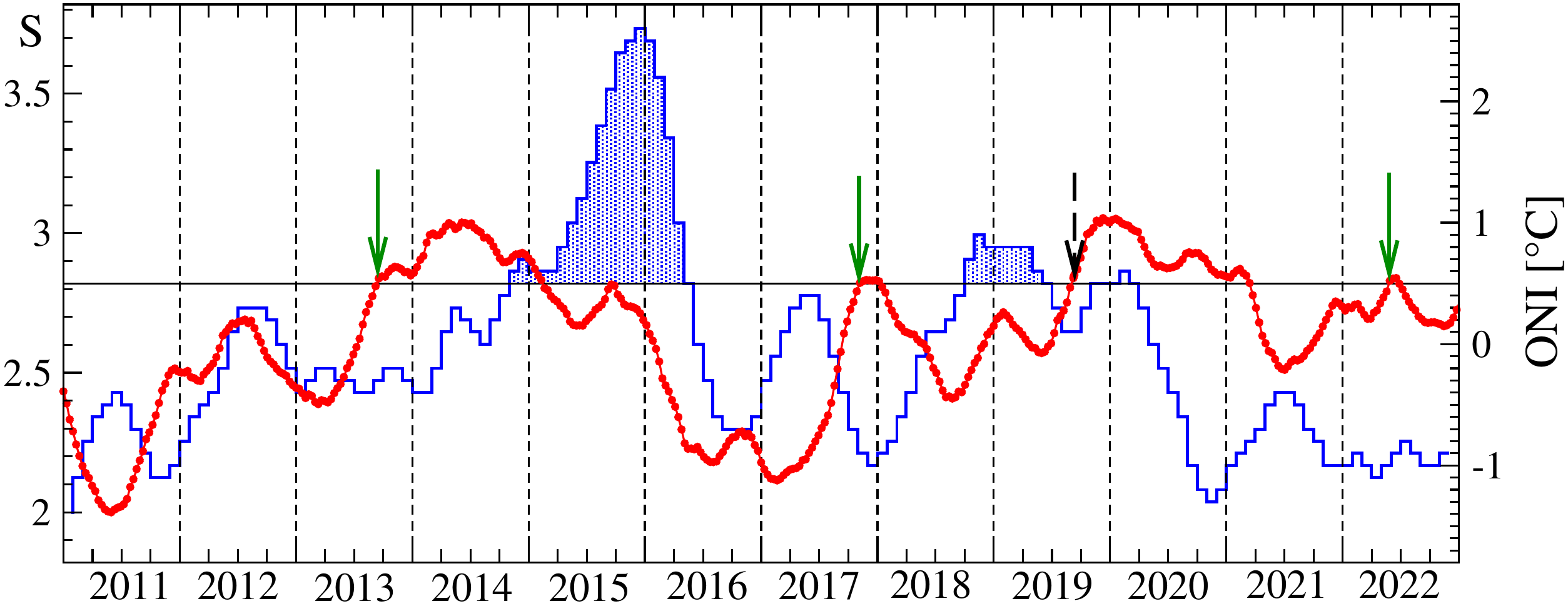}
\caption{ 
The climate network-based forecasting phase. Same as Fig. 2 but for the period between January 2011 and December 2022. Between May 24th and June 7th, 2022, the average link strength $S(t)$ crossed the critical threshold band from below, while the last ONI value was below 0.5°C (-1.1°C in MAM). Thus the method forecasted the onset of an El~Ni\~no in 2023. In the more restrictive version (ii) of the network-based algorithm, the false alarm in 2019 is not activated since the ONI increased to 0.5°C by the end of the year. In 2023, the ONI remained below 0.5°C throughout the year, thus the alarm is also activated in
version (ii) of the algorithm.
}
\label{fig3}
\end{center}
\end{figure}

\section{System Sample Entropy-based forecast}

\subsection{SysSampEn} 
The SysSampEn was introduced in \cite{Meng2019} as an analysis tool to quantify the complexity (disorder) in a complex system, in particular, in the temperature anomaly time series in the Ni\~no3.4 region.
It is approximately equal to the negative natural logarithm of the conditional probability that 2 subsequences similar (within a certain tolerance range) for $m$ consecutive data points remain similar for the next $p$ points, where the subsequences can originate from either the same or different time series (e.g., black
curves in Fig. 4), that is,
\begin{equation}
SysSampEn(m, p, l_{eff}, \gamma) = −log(\frac{A}{B}),
\end{equation}
where A is the number of pairs of similar subsequences of length $m + p$, $B$ is the number of pairs of similar subsequences of length
$m$, $l_{eff} \leq l$ is the number of data points used in the calculation for each time series of length $l$, and $\gamma$ is a constant which determines the tolerance range. The detailed definition of the SysSampEn for a general complex system composed of $N$ time series and how to objectively choose the parameter values is described in detail in \cite{Meng2019}.

\begin{figure}[]
\begin{center}
\includegraphics[width=9.5cm]{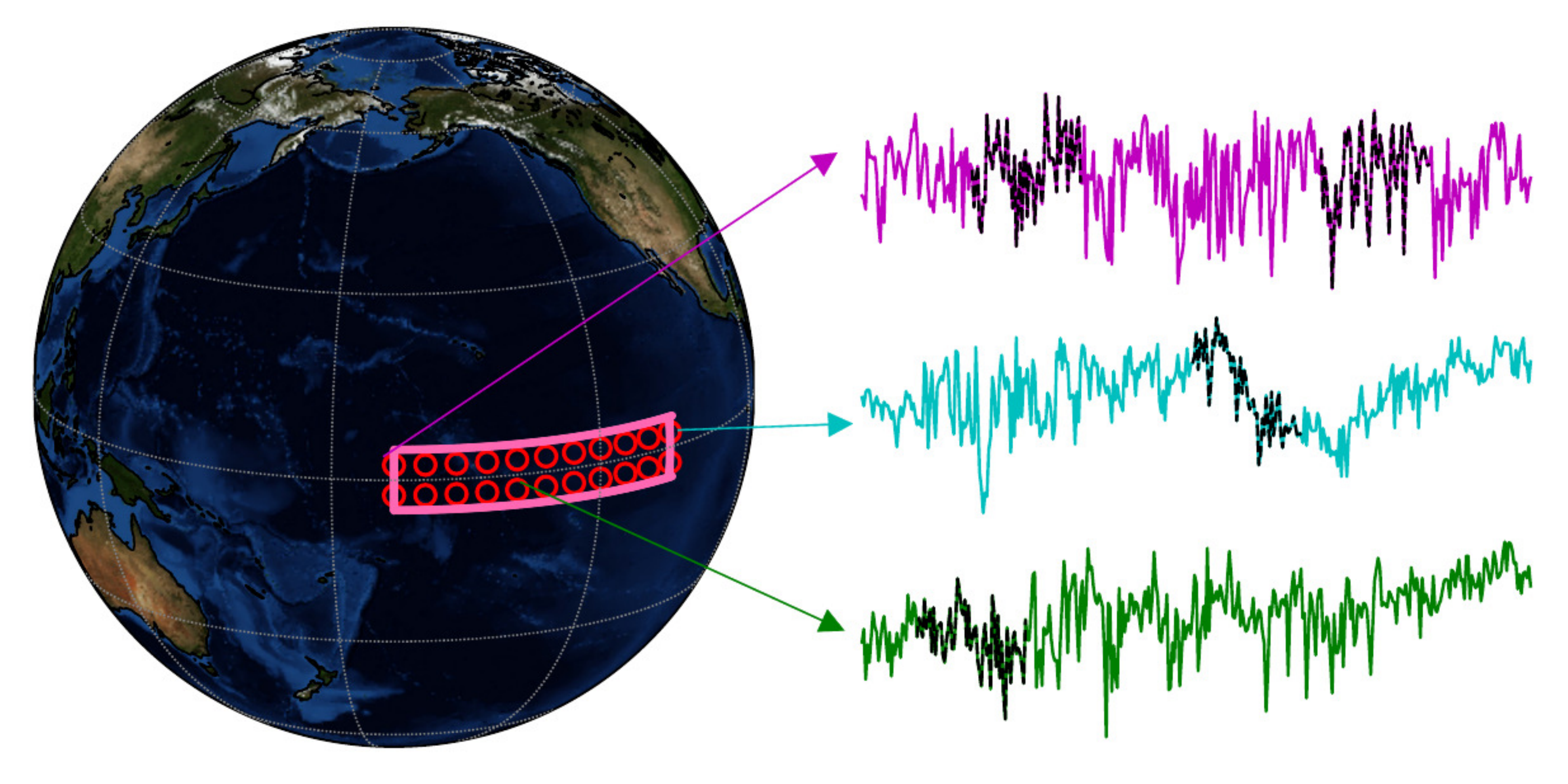}
\caption{The Ni\~no3.4 area and the SysSampEn input data.
The red circles indicate the 22 nodes covering the Ni\~no 3.4 region at a spatial resolution of $5^\circ \times 5^\circ$.
The curves are examples of the temperature anomaly time series for 3 nodes in the Ni\~no 3.4 region for one specific year.
Several examples of their subsequences are marked in black. In the calculation of the SysSampEn, both the similarity of subsequences within a time series and the similarity of subsequences of different time series are regarded.
Figure from \cite{Meng2019}.}
\label{fig4}
\end{center}
\end{figure}

In \cite{Meng2019}, it was found the previous year's ($y-1$) SysSampEn exhibits a strong positive correlation ($r=0.90$ on average) with the magnitude of an El Ni\~no in year $y$ when parameter combinations are used that are able to quantify a system's complexity with high accuracy. This linear relationship between SysSampEn and El Ni\~no magnitude enables thus to predict the magnitude of an upcoming El Ni\~no when the current ($y-1$) SysSampEn is inserted into the linear regression equation between the two quantities.

If the forecasted El~Ni\~no magnitude is below $0.5^\circ$C then the absence of an El Ni\~no onset is predicted for the following year $y$. Thus SysSampEn values below a certain threshold indicate the absence of an El Ni\~no onset. In contrast, if the SysSampEn is above this threshold and the ONI in December of the current year is below $0.5^\circ$C then the method predicts the onset of an El~Ni\~no event in the following year.

The SysSampEn represents a generalization of two information entropies that are widely used tools in physiological fields: the sample entropy (SampEn) and the Cross-SampEn \cite{Richman2000}. For details about their relation to the SysSampEn, see \cite{Meng2019}.

\subsection{Forecast for 2023}

Here we use as input data the daily near-surface (1000 hPa) air temperatures of the ERA5 reanalysis from the European Centre for Medium-Range Weather Forecasts (ECMWF) \cite{ERA5} analysed at a $5^\circ$  \ resolution. The last months in 2022 are from the initial data release ERA5T, which in contrast to ERA5, only lags a few days behind real-time.

The daily time series are preprocessed by subtracting the corresponding climatological mean and then dividing by the climatological standard deviation. We start in 1984 and use the previous years to calculate the first anomalies.
For the calculation of the climatological mean and standard deviation, only past data up to the year of the prediction are used. For simplicity, leap days are excluded. We use the same parameter values for the SysSampEn as in \cite{Meng2019}, $m = 30$, $p = 30$, $\gamma=8$ and $l_{eff} = 330$.

Figure 5 shows the results of the analysis. The magnitude forecast is shown as the height of rectangles in the year when the forecast is made, i.e., 1 year ahead of a potential El~Ni\~no onset. The forecast is obtained by inserting the regarded calendar year's SysSampEn value into the linear regression function between SysSampEn and El~Ni\~no magnitude. For the 2023 forecast, the regression is based on all correctly hindcasted El~Ni\~no events before 2022. The red curve shows the ONI and the red shades indicate the El~Ni\~no periods. The blue rectangles show the correct prediction of an El~Ni\~no in the following calendar year and grey rectangles with a violet border show false alarms.

There are 15 occurrences of high SysSampEn accompanied by a lower than $0.5^\circ$C ONI in December. In 9 of these cases, the hindcast was correct. White dashed rectangles show correct forecasts for the absence of an El~Ni\~no. 9 of the 10 El~Ni\~no events were preceded by a year with a high SysSampEn, the only missed event is the 2009/10 El~Ni\~no, where the preceding SysSampEn value was slightly below the threshold.

The forecasted El~Ni\~no magnitude for 2023 is $1.49\pm0.37^\circ$C, that is, well above $0.5^\circ$C, as shown by the green rectangle. The SysSampEn value for 2022 is $1.78$, i.e., well above the threshold value of $1.235$.
Figure \ref{fig6} shows the strong linear relationship (r=0.82) between the SysSampEn in the year before an El~Ni\~no onset and the maximal magnitude of the following El~Ni\~no, on which the prediction is based. The forecast's root-mean-square error (RMSE) ($\pm0.37^\circ$C) is obtained by leave-on-out hindcasting applied to the past El Niño events.

\begin{figure}[]
\begin{center}
\includegraphics[width=15cm]{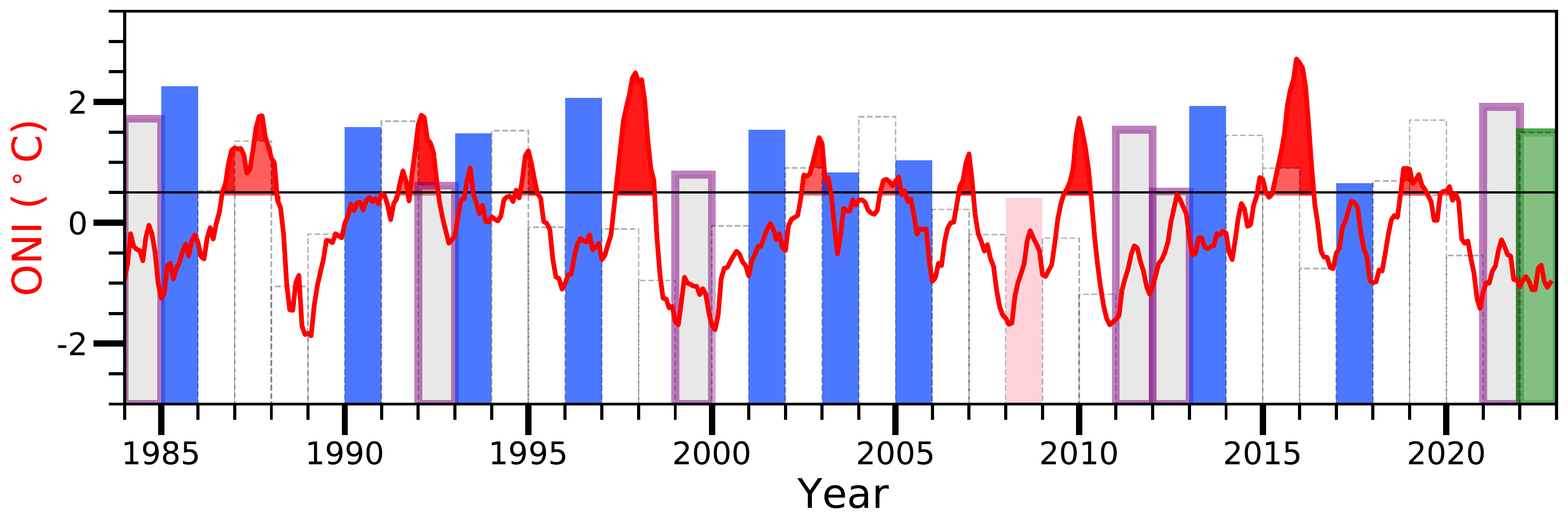}
\caption{Forecasted and observed El~Ni\~no magnitudes. The magnitude forecast is shown as the height of rectangles in the year when the forecast is made, i.e., one year ahead of a potential El~Ni\~no. The forecast is obtained by inserting the regarded calendar year's SysSampEn value into the linear regression function between SysSampEn and El~Ni\~no magnitude. To forecast the following year's condition, we use the ERA5 daily near-surface (1000 hPa) temperatures with the set of SysSampEn parameters ($m = 30$, $p = 30$, $\gamma=8$ and $l_{eff} = 330$). The red curve shows the ONI and the red shades highlight the El~Ni\~no periods.
The blue rectangles show the correct prediction of an El~Ni\~no in the following calendar year. 
The onset of an El~Ni\~no in the following year is predicted if the forecasted magnitude is above $0.5^\circ$C and the current year's December ONI is $<0.5^\circ$C. White dashed rectangles show correct forecasts for the absence of an El~Ni\~no.
Grey bars with a violet border show false alarms and the only missed event is shown as a pink rectangle.
For 2022 the SysSampEn has a high value of $1.78$, which results in a forecasted El~Ni\~no magnitude of $1.49\pm0.37$\textdegree C.
}
\label{fig5}
\end{center}
\end{figure}

\begin{figure}[]
\begin{center}
\includegraphics[width=9cm]{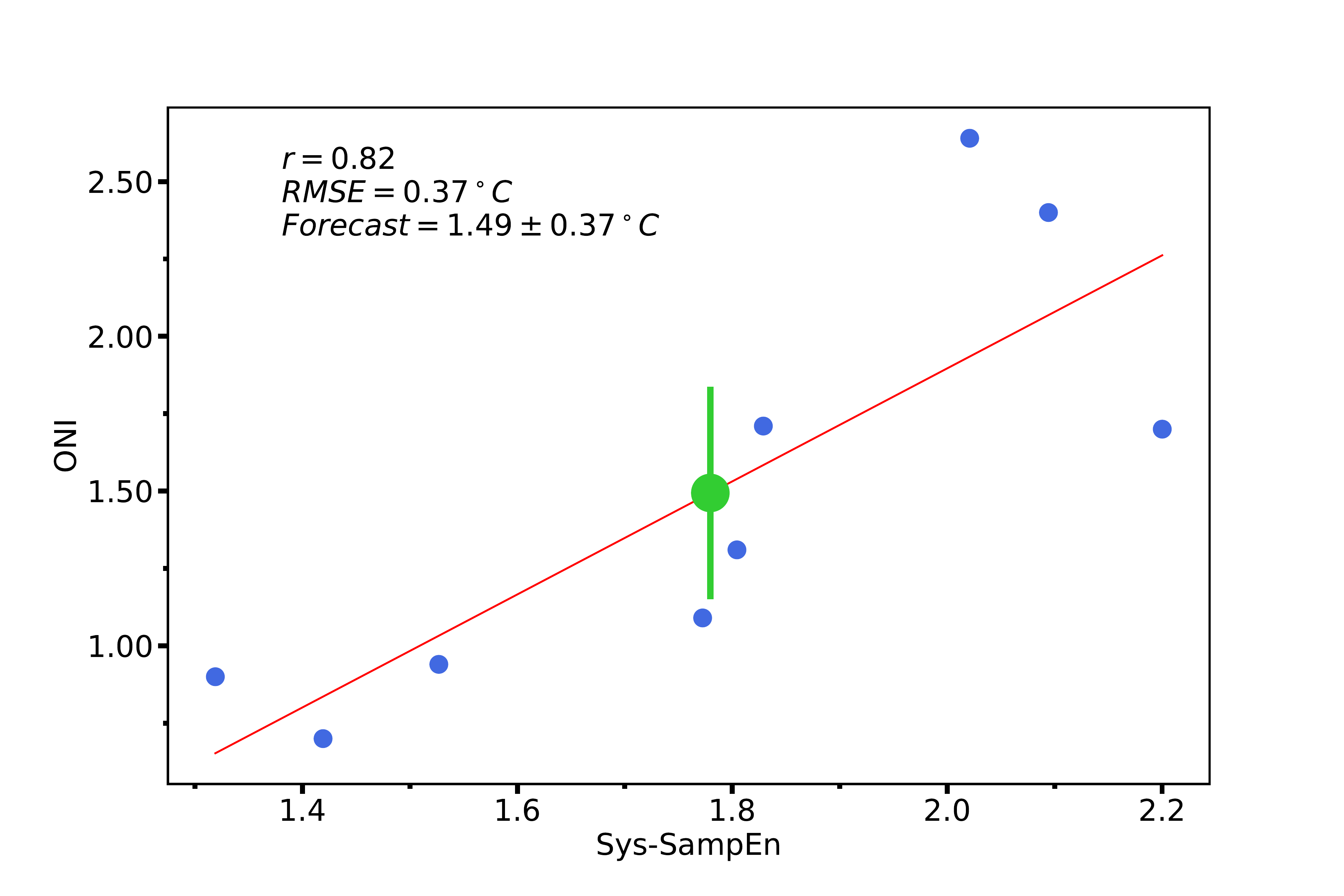}
\caption{Scatter plot of maximal El~Ni\~no magnitude versus the previous calendar year's SysSampEn (blue points). The red line shows the corresponding least-square linear fit. The green dot shows the magnitude forecast for 2023, based on the SysSampEn in 2022. The RMSE of the forecast (green vertical line) is obtained by leave-one-out hindcasting applied to the past El~Ni\~no events.
}
\label{fig6}
\end{center}
\end{figure}

\section{Implications for global temperature}
El~Ni\~no episodes lead, besides their local and regional weather impacts, also to a temporary increase of the global temperature with a time lag of about 2-5 months \cite{Trenberth2002, Foster2011}. At the end of 2021, the ONI was at $-1.0^\circ$C (OND), and also throughout 2022 it remained around $-1.0^\circ$C \cite{NOAA}. An El~Ni\~no episode in 2023 with a magnitude of $1.49\pm0.37^\circ$C (RMSE) would thus increase the ONI compared to 2022 by $2.49\pm0.37^\circ$C. If we assume a regression coefficient of $0.1^\circ$C per $1^\circ$C change in the Ni\~no3.4 area \cite{Trenberth2002,Foster2011, Trenberth2013, Berkley2022}, this corresponds to an increase in the monthly global temperatures of up to $0.249\pm0.037^\circ$C. The World Meteorological Organization (WMO) estimated the global temperature in 2022 compared to pre-industrial levels (1850-1900) at $+1.15^\circ$C [1.02 to 1.27](5\% to 95\% confidence interval) \cite{WMO2023, WMO2023b}. A list of global temperature estimates for 2022 and their reference periods are shown in Table 1.

A rough estimate of an El~Ni\~no, as forecasted here, on the global monthly temperatures in 2024 can thus be obtained by assuming a warming trend of $0.2^\circ$/decade: $1.15^\circ C + 2 * 0.02^\circ C + 0.249\pm0.037^\circ C = 1.439\pm0.037^\circ C$ (RMSE).
Assuming a Gaussian distribution for the uncertainty of the WMO estimation, this means that during 2024 the margin before breaching $1.5^\circ$C shrinks to less than 1 standard deviation (0.80*sd). This corresponds to a 21.1\% chance that $1.5^\circ$C will have been temporarily breached by the end of 2024.

The onset of the here forecasted El~Ni\~no would make a temporary breaching of $1.5^\circ$C in the Berkley Earth monthly data even likely. For this dataset, the 2022 estimate for the global temperature is $1.24\pm0.03^\circ$C. Based on this, an El~Ni\~no of the here forecasted magnitude would push the monthly global temperatures in 2024 up to $1.24^\circ C + 2 * 0.02^\circ C + 0.249\pm0.037^\circ C = 1.529\pm0.037^\circ C$. Since a moderate-to-strong El~Ni\~no might last up to the early boreal summer, the Berkeley Earth annual global temperature could also breach $+1.5^\circ$C (compared to pre-industrial levels) already in 2024.

\begin{table}
 \begin{center}
\renewcommand{\arraystretch}{1.1}
  \begin{tabular}{ l | c | r }
    \hline
    Source & 2022 estimate & baseline \\ \hline \hline
    Berkeley Earth & $+1.24\pm0.03^\circ$C & 1850-1900 \\ \hline
    ERA5 & $+1.18^\circ$C & 1850-1900 \\ \hline
    GISSTEMP & $+0.89^\circ$C & 1951-1980 \\ \hline
    HadCRUT5 & $+1.16\pm0.08^\circ$C & 1850-1900 \\ \hline
    NOAAGlobalTemp & $+0.86^\circ$C & 1901-2000 \\ \hline
    JRA-55 & -* & - \\ \hline \hline
    WMO  & $+1.15^\circ$C [1.02 to 1.27] & 1850-1900 \\
    \hline
  \end{tabular}
      \caption{Estimates of the global temperature in 2022 compared to the corresponding baseline periods \cite{Berkley2022,ERA5_2022,giss2022,HadCRUT5,NOAAGlobalTemp, WMO2023, WMO2023b}. (*Estimate not available at time of writing)}
\end{center}
\end{table}

\section*{Acknowledgements}
J.L. thanks the “Brazil East Africa Peru India Climate Capacities (B-EPICC)” project, which is part of  the International Climate Initiative (IKI) of the German Federal Ministry for Economic Affairs and Climate Action (BMWK) and implemented by the Federal Foreign Office (AA). J.M. and J.F. acknowledge the support by the National Natural Science Foundation of China (Grant No. 12205025, 12275020, 12135003).

\end{document}